\documentclass[aps,preprint,showpacs,showkeys]{revtex4}
\usepackage{graphicx}
\usepackage{amssymb}
\usepackage{bm}
\usepackage[all]{xy}
\usepackage{tikz}
\usepackage{amsmath}

\begin{document}
\setcounter{page}{1}
\title[]{Quantum Mechanics on a Poincar\'e Hyperboloid}
\author{HyunCheol \surname{Song}}
\author{Sang Gyu \surname{Jo}}
\email{sgjo@knu.ac.kr}
\affiliation{Department of Physics, Kyungpook National University, Daegu}
\date[]{Received }

\begin{abstract}
We discuss the process to obtain Poisson brackets among the phase-space variables of a system of a charged particle on a Poincar\'e hyperboloid in the presence of a uniform magnetic field. We show that after quantization the Dirac bracket algebra becomes the algebra of ISO(1,2). The representation of this algebra is explicitly analyzed and the Hamiltonian of this system has been derived.

\end{abstract}



\maketitle

\section{INTRODUCTION}

A Poincar\'e hyperboloid, as a mathematical object, has many peculiarities appealing to physicists' interests. It is a maximally symmetric and curved two-dimensional space. Unlike a sphere, which is also a maximally symmetric and curved two-dimensional space, a Poincar\'e hyperboloid is an open space with a negative scalar curvature. A Poincar\'e hyperboloid can be pictured as an embedded manifold in a three-dimensional Minkowski space with the embedding equation $x^{2}+y^{2}-z^{2}+a^{2}=0$. This embedded manifold is endowed with the induced metric which gives a negative Ricci scalar. Another aspect of a Poincar\'e hyperboloid is that all the Riemann surfaces with genus equal to or greater than two can be obtained by suitable identifications of points of this Poincar\'e hyperboloid\cite{Balazs}.

In this work, we analyze the quantum structure of a Poincar\'e hyperboloid. Starting from Poisson brackets among the phase-space variables of three-dimensional Minkowski space and introducing the constraint $x^{2}+y^{2}-z^{2} + a^{2}=0$, we derive the modified Poisson brackets among the phase-space variables of a classical system of a charged particle on a Poincar\'e hyperboloid on which a uniform magnetic field is applied. From these modified Poisson brackets we derive the Dirac brackets and show that the Dirac bracket algebra turns out to be the algebra of $ ISO(1, 2) $, the Poincar\'e group on $3$-dimensional Minkowski space. This algebra has 6 generators $\{\hat{x}, \hat{y}, \hat{z}, \hat{J}_{1}, \hat{J}_{2}, \hat{J}_{3}\}$. There are also two constraints $\hat{x}^{2}+\hat{y}^{2}-\hat{z}^{2}=-a^{2}$ and $\hat{x}\hat{J}_{1} + \hat{y}\hat{J}_{2} + \hat{z}\hat{J}_{3} = \frac{qg}{c}a$, both of which are the Casimir operators of this algebra\cite{Jakiw, Gitman, Binegar}. Here $q$ is the charge of the particle, $g$ is the parameter to describe the strength of the magnetic field and $c$ is the speed of light. (We take the CGS unit system.) These two constraints reduce the dimension of the phase-space from six to four which is the dimensionality of the phase-space of the Poincar\'e hyperboloid.

We also give the representation of this algebra. Due to the constraints, the Hilbert space is spanned by the states with quantum numbers corresponding to the coordinates of the Poincar\'e hyperboloid. We explicitly derive the matrix elemnets of the generators and the form of the Hamiltonian in this representation. In the special case where the magnetic field is absent ($g=0$), our result agrees with the known result\cite{Balazs, Davis, Argyres}.

In the next section, we consider a classical system of a charged particle on a Poincar\'e hyperboloid with a uniform magnetic field. In section III, we derive the Dirac bracket algebra and the representation of this algebra. Finally, in section IV, we give a brief summary.

\section{Classical Mechanics on a hyperboloid with Magnetic Field}

We begin with a 3-dimensional Minkowski space $M^{3}$ with metric $g_{ij} = \mathrm{diag}(1,1,-1)$. The coordinates of this space are $(x, y, z)$ and the metric is given by
\begin{eqnarray}
	ds^{2} = dx^{2} + dy^{2} -dz^{2} = dx^{i} dx_{i}. 
\end{eqnarray}
We are mainly concerned with the positive $z$-like region $Z^{+}$ defined by $Z^{+} = \{(x,y,z)|z>0, x^{2}+y^{2}-z^{2}<0\}$.

In $Z^{+}$, a magnetic field 
\begin{eqnarray}
	B^{i} &=& g \frac{x^{i}}{(- x\cdot x)^{\frac{3}{2}}} \label{magnetic_field}
\end{eqnarray}
is assumed to exist. Note that this magnetic field is well defined in $Z^{+}$ and invariant under the Lorentz transformation $SO(1,2)$. This field satisfies the divergence-free condition $\partial_{i}B^{i} = 0$ in $Z^{+}$. 
This magnetic field is the analog of the magnetic field generated by a magnetic monopole at the origin of a Euclidean space $R^{3}$.

Consider a particle of mass $m$ and electric charge $q$ moving in this space. The dynamics of the particle would be governed by the following Poisson brackets among the phase-space variables of this theory
\begin{eqnarray}
	\{x^{i},x^{j}\} &=& 0, \\
	\{x^{i},\pi_{j}\} &=& \delta^{i}_{j}, \\
	\{\pi_{i},\pi_{j}\} &=& \frac{q}{c} \epsilon_{ijk} B^{k}, 
\end{eqnarray}
and the Hamiltonian
\begin{eqnarray}
	H = \frac{1}{2m} \pi_{i} \pi_{j} g^{ij},
\end{eqnarray}
where $\epsilon^{ijk}$ is the totally antisymetric Levi-Civita symbol with $\epsilon^{123}=-1$ ($\epsilon_{123}=1$) and $g^{ij}$ is the inverse metric tensor. Note that the indices can be lowered and raised using the metric or the inverse metric tensor.

The Hamiltonian equations of motion come from 
\begin{eqnarray}
	\dot{x}^{i} &=& \{{x}^{i}, H \}, \\
	\dot{\pi}_{i} &=& \{{\pi}_{i}, H \}, 
\end{eqnarray}
and they are
\begin{eqnarray}
	{\pi}^{i} &=& m \dot{x}^{i} , \\
	m \ddot{x}^{i} &=&  \frac{q}{c}\epsilon^{ijk} \dot{x}_{j} B_{k}. \nonumber
\end{eqnarray}
The Hamiltonian in terms of velocity variables is  
\begin{eqnarray}
	H = \frac{m}{2}(\dot{x}^{2}+\dot{y}^{2}-\dot{z}^{2}).
\end{eqnarray}
This Hamiltonian has no lower bound and quantum version would be problematic.

We now introduce a constraint given by
\begin{eqnarray}
	C_{1} = x^{2} + y^{2}  - z^{2} + a^{2} = 0 . \label{C0}
\end{eqnarray}
The subspace satisfying this constraint is composed of two separate components. We choose the component with $z>0$(FIG. 1).
\begin{figure}[h]
	\centering
    \includegraphics[angle=0, width=0.5 \textwidth]{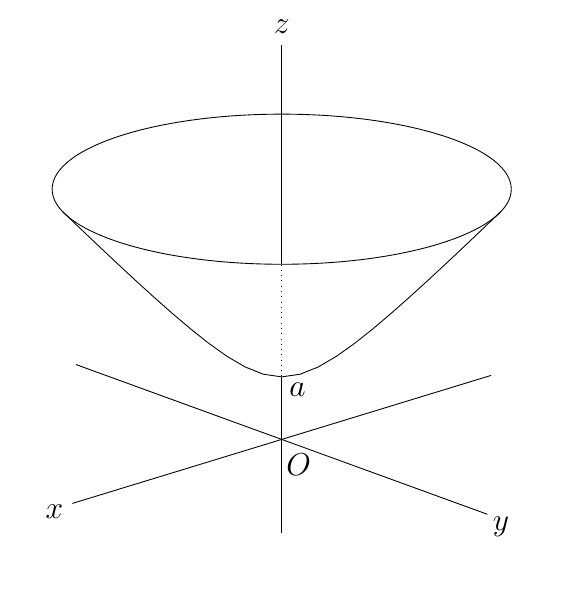}
    \caption{Poincar\'e Hyperboloid}
\end{figure}
This space is called Poincar\'e hyperboloid. We choose $(\theta, \phi)$ as the coordinates on this hyperboloid.
\begin{eqnarray}
& &    x = a\cos\phi \sinh\theta, \label{xxx} \\
& &    y = a\sin\phi \sinh\theta, \label{yyy} \\
& &    z = a\cosh\theta. \label{zzz}
\end{eqnarray}
The induced metric is given by
\begin{eqnarray}
    ds_{H}^{2} = a^{2} \sinh^{2} \theta d\phi^{2} + a^{2} d\theta^{2}.
\end{eqnarray}
This metric is positive definite and the subspace is spacelike. Later we use either $\vec{x}$ or $(\theta, \varphi)$ as a point on the hyperboloid.

To derive the Poisson brackets among the phase-space variables of the hyperboloid, we follow the standard procedure developed by Dirac\cite{Dirac, git tyu, Weinberg1, Weinberg2}. First we modify the Hamiltonian as
\begin{eqnarray}
	H_{1} = H_{0} + u_{1} C_{1}, 
\end{eqnarray}
with
\begin{eqnarray}
	H_{0} = \frac{1}{2m} \pi^{i} \pi_{i},
\end{eqnarray}
and $C_{1}$ being the primary constraint gievn in eq.(\ref{C0}). The secondary constraints are obtained by requiring the time derivative of existing constraints to vanish. Using the previous Poisson bracket relations, 
\begin{eqnarray}
	\dot{C}_{1} &=& \{C_{1}, H_{1}\}  \\
	            &=& \frac{2}{m}  x^{i} \pi_{i}, 
\end{eqnarray}
and this should vanish. So we have a secondary constraint $C_{2}$ defined by
\begin{eqnarray}
	C_{2} = x^{i}\pi_{i} = 0.
\end{eqnarray}
With this new constraint, we redefine the Hamiltonian as
\begin{eqnarray}
	H_{2} = H_{0} + u_{1} C_{1} + u_{2} C_{2},  \label{h222}
\end{eqnarray}
and we get
\begin{eqnarray}
	\dot{C}_{2} &=& \{C_{2}, H_{2}\}  \\
			   &=& \frac{1}{m}\pi^{i} \pi_{i} + 2 u_{1} (a^{2}-C_{1}).  
\end{eqnarray}
Choosing $u_{1}$ as 
\begin{eqnarray}
	u_{1} = - \frac{\pi^{i}\pi_{i}}{2ma^{2}} , 
\end{eqnarray}
we obtain
\begin{eqnarray}
	\dot{C}_{2} = - 2 u_{1} C_{1},
\end{eqnarray}
which vanishes automatically due to the primary constraint $C_{1}$. Therefore, we get only one secondary constraint which is $C_{2}$ and we end up with two constraints $C_{1}$ and $C_{2}$.

Defining 
\begin{eqnarray}
	M_{ij} = \{ C_{i}, C_{j} \},
\end{eqnarray}
we get 
\begin{eqnarray}
	M &=& \left(  \begin{array}{cc}
		     	                0 & -2 a^{2} \\
				                2 a^{2} & 0 
		               \end{array} \right),
\end{eqnarray}
and its inverse
\begin{eqnarray}
	M^{-1} &=& \frac{1}{2 a^{2}} \left( \begin{array}{cc}
		     	                0 & 1 \\
				                -1 & 0 
		               \end{array} \right).
\end{eqnarray}
We see that $C_{1}$ and $C_{2}$ are the second class constraints. As was suggested by Dirac\cite{Dirac}, we modify the Poisson brackets as 
\begin{eqnarray}
	\{A, B \}_{M} = \{A, B \} - \{A, C_{i} \} M^{-1}_{ij} \{ C_{j}, B \},
\end{eqnarray}
and we obtain
\begin{eqnarray}
	\{x^{i}, x^{j} \}_{M} 
					    &=& 0, 
\end{eqnarray}
\begin{eqnarray}
	\{x^{i}, \pi_{j} \}_{M} 
					      &=& \delta^{i}_{j} + \frac{1}{a^{2}} x^{i} x_{j},  
\end{eqnarray}
\begin{eqnarray}
	\{\pi_{i}, \pi_{j} \}_{M} 
					        &=& \frac{q}{c}\epsilon_{ijk} B^{k} + \frac{1}{a^{2}} (x_{i} \pi_{j} - \pi_{i} x_{j}). \label{magfield}
\end{eqnarray}
We now safely set $C_{1}$ and $C_{2}$ to be zero and the magnetic field in eq.(\ref{magnetic_field})  becomes  
\begin{eqnarray}
	B^{i} &=& g \frac{x^{i}}{a^{3}}. 
\end{eqnarray}
This magnetic field is orthogonal to the hyperboloid because 
\begin{eqnarray}
	B^{i} dx_{i} = \frac{g}{a^{3}} x^{i} dx_{i} = \frac{g}{2a^{3}} d(x^{i}x_{i})=0 ,
\end{eqnarray}
with $dx^{i}$ being an arbitrary displacement on the hyperboloid. Furthermore, its magnitude is a constant on the hyperboloid because 
\begin{eqnarray}
	B^{i}B_{i} = -\frac{g^{2}}{a^{4}}. 
\end{eqnarray}
In other words, this magnetic field is uniform on the hyperboloid. Substituting this $B^{i}$ into eq.(\ref{magfield}), we get
\begin{eqnarray}
	\{\pi_{i}, \pi_{j} \}_{M} 
					        &=& \frac{qg}{ca^{3}} \epsilon_{ijk} x^{k} + \frac{1}{a^{2}}(x_{i}\pi_{j} - x_{j}\pi_{i}).
\end{eqnarray}
The Hamiltonian in eq.(\ref{h222}) becomes 
\begin{eqnarray}
	H = \frac{1}{2m} \pi^{i} \pi_{i},
\end{eqnarray}
with two constraints 
\begin{eqnarray}
  C_{1} &=& x^{i} x_{i} + a^{2} = 0, \\
  C_{2} &=& x^{i} \pi_{i} = 0.   \label{c2xp123}
\end{eqnarray}

Now we analyze the dynamics. The equations of motion are  
\begin{eqnarray}
	\dot{x}^{i} &=& \{x^{i}, H\}_{M} \\
			   &=& \frac{1}{m} \pi^{i} ,
\end{eqnarray}
and
\begin{eqnarray}
	\dot{\pi}_{i} &=& \{\pi_{i}, H\}_{M} \\
			   &=& \frac{qg}{mca^{3}} \epsilon_{ijk} \pi^{j} x^{k} + \frac{1}{ma^{2}} x_{i} \pi^{j} \pi_{j}.
\end{eqnarray}
From these equations, we can show that $\pi^{i}\pi_{i}$ and $J_{i}$ defined below are constants of motion. 
\begin{eqnarray}
	J_{i} &\equiv& \epsilon_{ijk} x^{j} \pi^{k} - \frac{qg}{ca} x_{i}.  \label{jiequiv}
\end{eqnarray}
We also see that $x^{i}J_{i}$ is another constant of motion whose value is given by 
\begin{eqnarray}
	x^{i}J_{i} 
	           &=& \frac{qg}{c}a. \label{xijitil}
\end{eqnarray}
We set $\pi^{i}\pi_{i}$ to be $m^{2}v^{2}$ with $v = \sqrt{\dot{x}^{2}+\dot{y}^{2}-\dot{z}^{2}}$ being the speed of the particle on the hyperboloid. Then we get 
\begin{eqnarray}
	J^{i} J_{i} = m^{2} a^{2} v^{2} -\frac{q^{2} g^{2}}{c^{2}}.	   
\end{eqnarray}
The value of $J^{i}J_{i}$ can be any real number which is determined by the speed $v$. We will see that the sign of $J^{i}J_{i}$ determines the pattern of particle's motion. There are three cases; $J_{i}$ being $xy$-like, null, $z$-like
corresponding to the cases $J^{i}J_{i}$ being positive, zero, negative respectively. 
Defining $Q$ as 
\begin{eqnarray}
	Q=\frac{qg}{mcav}, 
\end{eqnarray}
we have the following correspondences;
\begin{eqnarray}
	J^{i} J_{i} \gtreqqless 0 \iff   |Q| \lesseqqgtr 1. 
\end{eqnarray}
In other words, $|Q|>1$\,($|Q|<1$) means $J_{i}$ is $xy$-like\,($z$-like) and $|Q|=1$ means $J_{i}$ is null. Note that $v=0$ is a solution where the particle stays at the initial position without any motion and here we assume $v \neq 0$. 
The orbit of particle's motion is determined by $J_{i}$ through eq.(\ref{xijitil}). 

In oder to visualize the orbit explicitly, we assume that the initial position at $t=0$ is $\vec{x}_{s}=(0,0,a)$ and that the initial velocity is $x$-directional; 
\begin{eqnarray}
	\dot{x}^{i}(t=0) &=& (v,0,0). 	
\end{eqnarray}  
Note that there is no loss of generality by this assumption. Any initial situation can be transformed to the assumed situation by a suitable Lorentz transformation. With the assumed initial conditions, we have 
\begin{eqnarray}
	(J_{i}) = (0, mav,\frac{qg}{c}).
\end{eqnarray}
Then, the orbit equation eq.(\ref{xijitil}) becomes
\begin{eqnarray}
	\frac{y}{Qa} + \frac{z}{a} = 1.  
\end{eqnarray}
This equation defines a plane parallel to the $x$-axis. Two points $(0,0,a)$ and $(0,Qa,0)$ are on this plane. The intersection of this plane with the hyperboloid is the trajectory of the particle(Fig. 2). 
The slope of the plane on the $yz$ plane is $(-\frac{1}{Q})$ and, therefore, the trajectory is closed only when $|Q|>1$. This is the case when $J_{i}$ is $z$-like and the trajectory is an ellipse. 
When $|Q|=1$, $J_{i}$ is null and the trajectory is parabolic. When $|Q|<1$, $J_{i}$ is $xy-$like and the trajectory is hyperbolic. If there is no magnetic field on the hyperboloid, then $g=0$ and $Q=0$. In this case, the trajectory is the intersection of the plane $y=0$ and the hyperboloid. This trajectory is a geodesic on the hyperboloid\cite{Balazs}. The appearence of magnetic field deviates the trajectory from a geodesic. 

In terms of the speed of the particle, we can summarize as follows. If the speed $v$ is greater(less) than $v_{c}$, then the trajectory is hyperbolic(elliptic). If the speed $v$ is equal to $v_{c}$, then the trajectory is parabolic. Here, $v_{c}$ is the critical speed. 
\begin{eqnarray}
	v_{c} =\frac{|qg|}{mca}.
\end{eqnarray}

\begin{figure}[h]
	\centering
    \includegraphics[angle=0, width=0.6\textwidth]{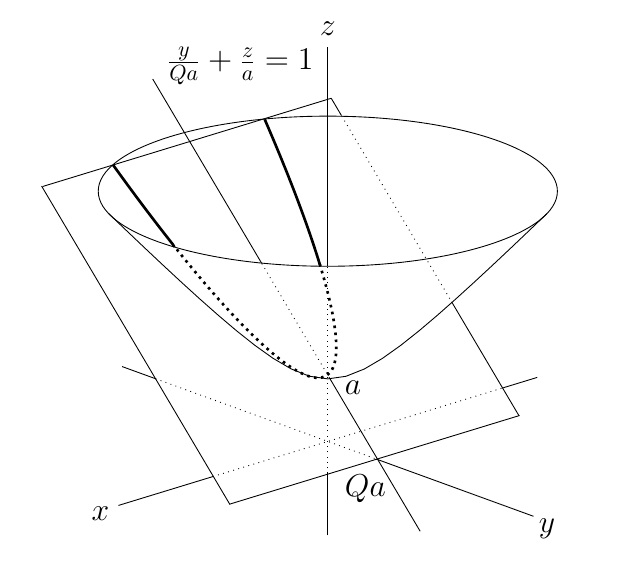}
    \caption{Trajectory}
\end{figure}

\section{Quantum Mechanics on a hyperboloid with Magnetic Field}

Before quantizing this theory, we summarize the Poisson bracket relations between new phase variables $(x^{i}, J^{i})$. 
\begin{eqnarray}
	\{x^{i}, x^{j} \}_{M} &=& 0, \label{starstar112}\\
	\{J^{i}, x^{j} \}_{M} &=& \epsilon^{ijk}x_{k} , \label{starstar113}\\
	\{J^{i}, J^{j} \}_{M} &=& \epsilon^{ijk}J_{k}.  \label{starstar114}
\end{eqnarray}
The constraints are 
\begin{eqnarray}
	C_{1} &=& x^{i} x_{i} + a^{2} = 0, \\
	C'_{2} &=& x^{i} J_{i} - \frac{qg}{c}a = 0 .
\end{eqnarray}
The constraint $C'_{2}$ comes from eq.(\ref{jiequiv}).
Note that the Poisson bracket relations between the old phase variables $(x^{i},\pi_{j})$ can be obtained from eq.(\ref{starstar112}-\ref{starstar114}) using the relation given by 
\begin{eqnarray}
	\pi_{i} = -\frac{1}{a^{2}} \epsilon_{ijk}x^{j}J^{k}. \label{pipi2}
\end{eqnarray}
When we go back to original phase variables $(x^{i},\pi_{j})$, the constraint $C_{2}$ in eq.(\ref{c2xp123}) comes from this equation(eq.(\ref{pipi2})). From now on we will work with new phase variables $(x^{i}, J_{j})$ and we drop the prime from $C_{2}'$ and use $C_{2}$ as 
\begin{eqnarray}
	C_{2} &=& x^{i} J_{i} - \frac{qg}{c}a = 0 .
\end{eqnarray}
The Hamiltonian in terms of new phase variables is
\begin{eqnarray}
	H = \frac{J^{i}J_{i}}{2ma^{2}}.
\end{eqnarray}

\subsection{Dirac Bracket Relations}

The transition from classical to quantum physics is achieved by changing Poisson brackets to Dirac brackets as $[A, B]=i\hbar\{A, B\}_{M}$. We get
\begin{eqnarray}
	\left[\hat{x}^{i}, \hat{x}^{j}\right] &=& 0, \label{comxx1} \\
	\left[\hat{J}^{i}, \hat{x}^{j}\right] &=& i\hbar\epsilon^{ijk}\hat{x}_{k},  \label{comjx2}\\
	\left[\hat{J}^{i}, \hat{J}^{j}\right] &=& i\hbar\epsilon^{ijk}\hat{J}_{k}.  \label{comjj3}
\end{eqnarray}
We assume that $(\hat{x}^{i}, \hat{J}^{j})$ are all hermitian. The constraints are
\begin{eqnarray}
	\hat{C_{1}} &=& \hat{x}^{i}\hat{x}_{i}+a^{2 }=0, \\
    \hat{C}_{2} &=& \hat{x}^{i} \hat{J}_{i} - \frac{qg}{c}a =0.
\end{eqnarray}
Two constraints are also hermitian. It can be checked that these constraints are compatible with the Dirac bracket relations. In other words,
\begin{eqnarray}
	\left[\hat{x}^{i}, \hat{C}_{(j)} \right] = \left[\hat{J}^{i}, \hat{C}_{(j)} \right]=0.
\end{eqnarray}
The Hamiltonian is
\begin{eqnarray}
	\hat{H} = \frac{\hat{J}^{i}\hat{J}_{i}}{2ma^{2}}.  \label{hamil1}
\end{eqnarray}
Note here that  momentum operator $\hat{\pi_{i}}$ is related to operators $\hat{x}^{i}, \hat{J}^{j}$ through
\begin{eqnarray}
	\hat{\pi}_{i} = -\frac{1}{a^{2}} \epsilon_{ijk} \frac{\hat{x}^{j} \hat{J}^{k}+\hat{J}^{k} \hat{x}^{j}}{2} =-\frac{1}{a^{2}}\epsilon_{ijk}\hat{x}^{j}\hat{J}^{k} - \frac{i\hbar}{a^{2}}\hat{x}_{i}. \label{equality}
\end{eqnarray}
The modification from eq.(\ref{pipi2}) is necessary to make $\hat{\pi}_{i}$ hermitian. The Dirac bracket relations among $(\hat{x}^{i}, \hat{\pi}_{j} )$ are as follows:
\begin{eqnarray}
	\left[\hat{x}^{i}, \hat{x}^{j}\right] &=& 0, \\
	\left[\hat{x}^{i}, \hat{\pi}_{j}\right] &=& i\hbar\delta^{i}_{j}+i\hbar\frac{1}{a^{2}}\hat{x}^{i}\hat{x}_{j}, \\
	\left[\hat{\pi}_{i}, \hat{\pi}_{j}\right] &=& i\hbar\left[\epsilon_{ijk}\frac{qg}{ca^{3}} \hat{x}^{k} + \frac{\hat{x}_{i}\hat{\pi}_{j} - \hat{x}_{j}\hat{\pi}_{i}}{a^{2}} \right].
\end{eqnarray}
From eq.(\ref{equality}), we get 
\begin{eqnarray}
	\hat{J}^{i} = \epsilon^{ijk}\hat{x}_{j}\hat{\pi}_{k} - \frac{qg}{ca}\hat{x}^{i}, 
\end{eqnarray}
which agrees with the calssical relation in eq.(\ref{jiequiv}).

We now observe that the algebra generated by $(\hat{x}^{i}, \hat{J}^{j})$ is that of the Poincar\'e group ($ISO(1,2)$). Note that $\hat{J}^{1}$($-\hat{J}^{2}$) generates the $y$-directional($x$-directional) boost and $\hat{J}^{3}$ generates the $z$-axis rotation. The Casimir operators of this algebra are $\hat{x}^{i}\hat{x}_{i}$ and $\hat{x}^{i}\hat{J}_{i}$, which in our case are required to be $(-a^{2})$ and $\frac{qg}{c}a $ respectively.

\subsection{Representation}

Following the procedure introduced in ref.\cite{Weinberg1}, we choose $\{\hat{x}^{i}\}$ as the generators of a maximal commuting subalgebra, and diagonalize them by the eigenkets $|\,\vec{x},\sigma>$ with $\sigma$ as extra quantum numbers to be fixed later;
\begin{eqnarray}
	\hat{x}^{i}|\,\vec{x}, \sigma> = x^{i} |\,\vec{x}, \sigma>.
\end{eqnarray}
Because of the constraint $\hat{x}_{i}\hat{x}^{i}+a^{2}=0$, the eigenvalue $x^{i}$ should satisfy ${x}_{i}{x}^{i}=-a^{2}$ and $\vec{x}$ corresponds to a point on the hyperboloid. The remaining three operators $\{\hat{J}^{i}\}$ generate the Lorentz group $SO(1,2)$ and this group acts on $|\,\vec{x},\sigma>$ in such a way that $\vec{x}$ undergoes the Lorentz transformation. In order to analyze how the extra quantum numbers $\sigma$ transform under the group action, we consider the eigen ket $|\vec{x}_{s},\sigma>(=|(0,0,a), \sigma>)$. We first note that the stability group of $\vec{x}_{s}$ is the group generated by $\hat{J}^{3}$. This subgroup, which is the group of the rotations around the $z$ axis, is one-dimensional. It is represented by one real quantum number $\sigma$. We, therefore, have
\begin{eqnarray}
	exp\left(i\frac{\theta \hat{J}^{3}}{\hbar}\right)|\vec{x}_{s}, \sigma> = exp\left(i\frac{\theta \sigma}{\hbar}\right)|\vec{x}_{s}, \sigma>,
\end{eqnarray}
or simply
\begin{eqnarray}
	\hat{J}^{3}|\vec{x}_{s}, \sigma> = \sigma |\vec{x}_{s}, \sigma>.
\end{eqnarray}
We now impose the constraint $\hat{x}^{j}\hat{J}_{j}=\frac{qg}{c}a$ to determine the value of $\sigma$.
\begin{eqnarray}
	\hat{x}^{j}\hat{J}_{j}|\vec{x}_{s}, \sigma> = a\hat{J}_{3}|\vec{x}_{s}, \sigma>= a\sigma|\vec{x}_{s}, \sigma>=\frac{qg}{c}a|\vec{x}_{s}, \sigma>.
\end{eqnarray}
Therefore, the value of $\sigma$ should be $\frac{qg}{c}$ and no other quantum number than $\vec{x}$ appears in the representation. The quantum number $\sigma$ is fixed to be $\frac{qg}{c}$ and we drop $\sigma$ from $|\vec{x}, \sigma>$. The representation space is spanned by $\{ |\,\vec{x}> \}$ with $\vec{x}=(x,y,z)$ on the hyperboloid and the coordinates $x, y, z$ are given in eq.(\ref{xxx}-\ref{zzz}). Later we sometimes use $|\theta,\varphi>$ instead of $|\vec{x}>$ using these coordinates.

We consider $\hat{U}(\Lambda)|\vec{x}>$ with $\Lambda$ being a Lorentz transformtion and $\hat{U}(\Lambda)$ being the unitary representation of the transformation. Applying $\hat{x}^{i}$ on this state, we get
\begin{eqnarray}
	\hat{x}^{i}\hat{U}(\Lambda)|\vec{x}> = \hat{U}(\Lambda) \hat{U}(\Lambda^{-1})\hat{x}^{i} \hat{U}(\Lambda) |\vec{x}> = (\Lambda\vec{x})^{i}\hat{U}(\Lambda)|\vec{x}>.
\end{eqnarray}
Here we used the identity
\begin{eqnarray}
	\hat{U}(\Lambda^{-1})\hat{x}^{i} \hat{U}(\Lambda) = \Lambda^{i}_{\,j} \hat{x}^{j}.
\end{eqnarray}
This indicates 
\begin{eqnarray}
	\hat{U}(\Lambda)|\vec{x}> = \mathrm{(phase)}|\Lambda\vec{x}>, \label{uuurep}
\end{eqnarray}
where (phase) is a phase factor depending on $\Lambda$ and $\vec{x}$. We define 
\begin{eqnarray}
	|\vec{x}> = \hat{U}(B(\vec{x}))|\vec{x}_{s}>,
\end{eqnarray}
where $B(\vec{x})$ is the pure boost which transforms $\vec{x}_{s}$ to $\vec{x}$. 
Then, eq.(\ref{uuurep}) becomes
\begin{eqnarray}
	\hat{U}(B^{-1}(\Lambda\vec{x})\, \Lambda \, B(\vec{x})) |\vec{x}_{s}> = \mathrm{(phase)}|\vec{x}_{s}>.
\end{eqnarray} 
Here we used $\hat{U}(g_{1})\hat{U}(g_{2})=\hat{U}(g_{1}g_{2})$.
Furthermore,
\begin{eqnarray}
	B^{-1}(\Lambda\vec{x})\, \Lambda \, B(\vec{x}) \vec{x}_{s} = \vec{x}_{s},
\end{eqnarray} 
which indicates 
\begin{eqnarray}
	B^{-1}(\Lambda\vec{x})\, \Lambda \, B(\vec{x}) = R(\phi(\Lambda,\vec{x})), \label{uuurep1}
\end{eqnarray} 
where $R(\phi)$ is the rotation by an angle $\phi$ around the $z$-axis and $\phi(\Lambda,\vec{x})$ is the angle  to be determined by the choice of $\Lambda$ and $\vec{x}$.
Therefore, the phase factor on the right hand side of eq.(\ref{uuurep}) should be $\mathrm{exp} (\frac{i\phi(\Lambda,\vec{x})}{\hbar}\sigma)$ and the equation becomes
\begin{eqnarray}
	\hat{U}(\Lambda)|\vec{x}>= e^{\frac{i\phi(\Lambda,\vec{x})}{\hbar}\sigma}|\Lambda\vec{x}>. \label{uuurep0}
\end{eqnarray} 
We parameterize $\Lambda$ as
\begin{eqnarray}
	\Lambda = R(\phi_{\Lambda})B(\theta_{\Lambda}, \varphi_{\Lambda}),
\end{eqnarray}
and we get 
\begin{eqnarray}
	\phi(\Lambda,\vec{x}) = \phi_{\Lambda} + 2 \tan^{-1} \left[\frac{\tanh\frac{\theta_{\Lambda}}{2}\tanh\frac{\theta_{x}}{2}\sin(\varphi_{\Lambda}-\varphi)}{1+\tanh\frac{\theta_{\Lambda}}{2}\tanh\frac{\theta_{x}}{2}\cos(\varphi_{\Lambda}-\varphi)}\right]. \label{uuurep11}
\end{eqnarray} 
The procedure to get $\phi(\Lambda,\vec{x})$ is given in the Appendix. In order to analyze $\hat{U}(\Lambda)$ action on a general state $|\psi>$, we introduce further structure of the Hilbert space spanned by the basis $\{|\vec{x}(\theta,\varphi)>|\vec{x}\in \mathrm{Hyperboloid}\}$. 
The inner product is given by
\begin{eqnarray}
	<\vec{x}(\theta,\varphi)|\vec{x}'(\theta',\varphi')> = \delta(\cosh\theta-\cosh\theta') \delta(\varphi-\varphi'),
\end{eqnarray}
and the completeness relation is 
\begin{eqnarray}
	I=\int d\mu |\vec{x}(\theta,\varphi)><\vec{x}'(\theta',\varphi')| , 
\end{eqnarray}
where $d\mu = \sinh\theta d\theta d\varphi$.

Consider a general state $|\psi>$ given by
\begin{eqnarray}
	|\psi>=\int d \mu \, \psi(\theta,\varphi)|\vec{x}(\theta,\varphi)>,
\end{eqnarray}
where $\psi(\theta,\varphi)=<\vec{x}(\theta,\varphi)|\psi>$. Applying $\hat{U}(\Lambda)$ on $|\psi>$ and using eq.(\ref{uuurep0}), we get
\begin{eqnarray}
	\hat{U}(\Lambda)|\psi>=\int d \mu \, \psi(\theta,\varphi)e^{\frac{i}{\hbar}\phi(\Lambda,\vec{x})\sigma}|\Lambda\vec{x}>.
\end{eqnarray}
Changing the variables of integration from $\vec{x}$ to $\vec{x}\,'=\Lambda\vec{x}$, we get 
\begin{eqnarray}
	\hat{U}(\Lambda)|\psi>=\int d \mu \, \psi(\Lambda^{-1}\vec{x})e^{\frac{i}{\hbar}\phi(\Lambda,\Lambda^{-1}\vec{x})\sigma}|\vec{x}>.
\end{eqnarray}
From this, we obtain
\begin{eqnarray}
	<\vec{x}|\hat{U}(\Lambda)|\psi>=e^{\frac{i}{\hbar}\phi(\Lambda,\Lambda^{-1}\vec{x})\sigma}\psi(\Lambda^{-1}\vec{x}). \label{xhatupsi}
\end{eqnarray}
Using eq.(\ref{uuurep11}), we can calculate $\phi(\Lambda,\Lambda^{-1}\vec{x})$. Detail of the calculation is given in the Appendix. The value is 
\begin{eqnarray}
	\phi(\Lambda,\Lambda^{-1}\vec{x}) = \phi_{\Lambda} + 2 \tan^{-1} \left[ \frac{\tanh\frac{\theta_{\Lambda}}{2}\tanh\frac{\theta}{2}\sin(\phi_{\Lambda}+\varphi_{\Lambda}-\varphi)}{1-\tanh\frac{\theta_{\Lambda}}{2}\tanh\frac{\theta}{2}\cos(\phi_{\Lambda}+\varphi_{\Lambda}-\varphi)} \right].
\end{eqnarray}

\subsection{Calculation of matrix elements.}

We now calculate the matrix elements of $\hat{J}^{i}$, $<\vec{x}|\hat{J}^{i}|\psi>$. We first take $i=2$. To calculate  $<\vec{x}|\hat{J}^{2}|\psi>$, we take $\Lambda$ to be an infinitesimal boost along $x$ direction
\begin{eqnarray}
	\Lambda &=& \Lambda(\phi_{\Lambda}=0, \theta_{\Lambda}=\epsilon,\varphi_{\Lambda}=0) \label{lambda123}\\
		    &\simeq& I + \epsilon \left(  \begin{array}{ccc}
		     	                     0    & 0 & 1 \\
				                     0    & 0 & 0 \\
				                     1    & 0 & 0 
		               \end{array} \right). 
\end{eqnarray}
Then, $\hat{U}(\Lambda)$ becomes
\begin{eqnarray}   
	\hat{U}(\Lambda) &=& e^{-\frac{i}{\hbar}\epsilon \hat{J}^{2}} \\
	                 &\simeq& I - \frac{i}{\hbar} \hat{J}^{2}. \label{phiphi1}
\end{eqnarray}
With this $\Lambda$, $\Lambda^{-1}\vec{x}$ becomes
\begin{eqnarray}
	\Lambda^{-1}\vec{x} = \left(  \begin{array}{c}
		     	                     x' \\
				                     y'  \\
				                     z' 
		               \end{array} \right) = \left(  \begin{array}{c}
		     	                     x - \epsilon z \\
				                     y  \\
				                     z - \epsilon x 
		               \end{array} \right).
\end{eqnarray}
Therefore,
\begin{eqnarray}
	<\Lambda^{-1}\vec{x}|\psi> = \left[1-\epsilon(z\frac{\partial}{\partial x}+x\frac{\partial}{\partial z}) \right] <\vec{x}|\psi>.   \label{phiphi2}
\end{eqnarray} 
We also have
\begin{eqnarray}
	\phi(\Lambda, \Lambda^{-1}\vec{x}) &=& 0 + 2\tan^{-1}\left[\frac{-\tanh\frac{\epsilon}{2} \tanh\frac{\theta}{2}\sin\varphi}{1-\tanh\frac{\epsilon}{2} \tanh\frac{\theta}{2} \cos\varphi }\right]   \\
	&\simeq& -\epsilon \tanh\frac{\theta}{2} \sin\varphi,
\end{eqnarray}
for $\Lambda$ being an infinitesimal $x$-directional boost given in eq.(\ref{lambda123}). Noting that $\tanh\frac{\theta}{2} \sin\varphi = \frac{y}{a+z}$, we have
\begin{eqnarray}
	\phi(\Lambda, \Lambda^{-1}\vec{x}) = -\epsilon \, \frac{y}{a+z}. \label{phiphi3}
\end{eqnarray}
Substituting  eq.(\ref{phiphi1}), (\ref{phiphi2}), (\ref{phiphi3}) into eq.(\ref{xhatupsi}), we get
\begin{eqnarray} 
	<\vec{x}\,|\hat{J}^{2}|\,\psi> = \left[ \frac{\hbar}{i} (z\frac{\partial}{\partial x} + x\frac{\partial}{\partial z}) 	+\frac{y}{a+z} \sigma \right] <\vec{x}|\psi> .
\end{eqnarray}
Taking $\Lambda$ to be an infinitesimal $y$-directional boost or an infinitesimal rotation, we get similar equations for $<\vec{x}|J^{1}|\psi>$ and $<\vec{x}|J^{3}|\psi>$ and they are
\begin{eqnarray} 
	<\vec{x}\,|\hat{J}^{1}|\,\psi> &=& \left[ -\frac{\hbar}{i} (z\frac{\partial}{\partial y} + y\frac{\partial}{\partial z}) 	+\frac{x}{a+z} \sigma \right] <\vec{x}|\psi> ,\\
	<\vec{x}\,|\hat{J}^{3}|\,\psi> &=& \left[ \frac{\hbar}{i} (y\frac{\partial}{\partial x} - x\frac{\partial}{\partial y}) 	+ \sigma \right] <\vec{x}|\psi>. \label{xj3psi}
\end{eqnarray}
The above three equations can be summarized into a single equation;
\begin{eqnarray} 
	<\vec{x}\,|\hat{J}^{i}|\,\psi> = \left[ \frac{\hbar}{i} \epsilon^{ijk} x_{j} \frac{\partial}{\partial x^{k}}  - \frac{a(x^{i}+x^{i}_{s})}{(x^{j}+x^{j}_{s})x_{j}} \sigma \right] <\vec{x}|\psi> ,
\end{eqnarray}
where $x^{i} = (x,y,z)$, $x_{i} = (x,y,-z)$ and $x^{i}_{s} = (0,0,a)$. 
In terms of angle variables, we get 
\begin{eqnarray} 
	<\theta,\varphi \,|\hat{J}^{1}|\,\psi> &=& \left[ -\frac{\hbar}{i} (\sin\varphi \frac{\partial}{\partial \theta} + \coth\theta\cos\varphi\frac{\partial}{\partial \varphi}) + \sigma \tanh\frac{\theta}{2}\cos\varphi \right] <\theta,\varphi|\psi> , \label{J11} \\
	<\theta,\varphi \,|\hat{J}^{2}|\,\psi> &=& \left[ \frac{\hbar}{i} (\cos\varphi \frac{\partial}{\partial \theta} - \coth\theta\sin\varphi\frac{\partial}{\partial \varphi})+ \sigma \tanh\frac{\theta}{2}\sin\varphi \right] <\theta,\varphi|\psi> , \label{J12}\\
	<\theta,\varphi \,|\hat{J}^{3}|\,\psi> &=& \left[ -\frac{\hbar}{i} \frac{\partial}{\partial \varphi} + \sigma \tanh\frac{\theta}{2}\cos\varphi \right] <\theta,\varphi|\psi> . \label{J13}
\end{eqnarray}
Note that $a$ does not appear in these equations. Taking $|\psi>$ to be $|\theta', \varphi'>$, we get $<\theta, \phi|\hat{J}^{i}|\theta', \phi'>$ from the above equation with $<\theta,\varphi|\psi>$ substituted by $\delta(\cosh\theta-\cosh\theta')\delta(\phi-\phi')$.

\subsection{Schr\"odinger equation}

In order to derive the Schr\"odinger equation, we take the matrix element of the Hamiltonian in eq.(\ref{hamil1});  
\begin{eqnarray} 
	<\theta,\varphi\,|\hat{H}|\,\psi> = \frac{1}{2ma^{2}} <\theta,\varphi|\hat{J}^{i}\hat{J}_{i} |\psi>. 
\end{eqnarray} 
Using eq.(\ref{J11}-\ref{J13}), the right hand side becomes
\begin{eqnarray} 
	\frac{1}{2ma^{2}} <\theta,\varphi| \hat{J}^{i} \hat{J}_{i} |\psi>  = - \frac{\hbar^{2}}{2ma^{2}} \left[ \frac{1}{\sinh\theta} \frac{\partial}{\partial\theta}\left(\sinh\theta\frac{\partial}{\partial\theta}\right) + \frac{1}{\sinh^{2}\theta}\frac{\partial^{2}}{\partial\phi^{2}}  \right. \nonumber \\
\left. + \frac{2}{1+\cosh\theta} \frac{\sigma^{2}}{\hbar^{2}} + i \frac{2\sigma}{(1+\cosh\theta)\hbar} \frac{\partial}{\partial\phi} \right]<\theta,\varphi|\psi>, 	 
\end{eqnarray}
and we get the Schr\"odinger equation;
\begin{eqnarray} 
	- \frac{\hbar^{2}}{2ma^{2}} \left[ \frac{1}{\sinh\theta} \frac{\partial}{\partial\theta}\left(\sinh\theta\frac{\partial}{\partial\theta}\right) + \frac{1}{\sinh^{2}\theta}\frac{\partial^{2}}{\partial\phi^{2}}  \right. \nonumber \\
\left. + \frac{2}{1+\cosh\theta} \frac{\sigma^{2}}{\hbar^{2}} + i \frac{2\sigma}{(1+\cosh\theta)\hbar} \frac{\partial}{\partial\phi} \right]\psi(\theta,\varphi) = E \psi (\theta,\varphi).    \label{Hamil0}
\end{eqnarray}    
If we take $\sigma=\frac{qg}{c}=0$, which corresponds to the case with no magnetic field, the above equation is the Schr\"odinger equation of the free theory on a hyperboloid previously discussed by several authors\cite{Balazs, Argyres}. We are working on the analysis of this Schr\"odinger equation with $\sigma \neq 0$.

\section{Summary and Discussion}

We derived the Poisson brackets among the phase-space variables for the system of a charged particle on a hyperboloid where a uniform magnetic field is applied. The hyperboloid is embedded in the 3-dimensional Minkowski space with $z$ as a time-like variable. The phase space variables are chosen to be $\{x^{i}, J_{j}\}$ with two constraints $x^{i}x_{i}=-a^{2}$ and $x^{i}J_{i}=\frac{qg}{c}a$. After quantization, we have shown that commutation relations among these variables given in eq.(\ref{comxx1}-\ref{comjj3}) form the algebra of ISO(1,2). The operators in the constraints, $\hat{x}^{i}\hat{x}_{i}$ and $\hat{x}^{i}\hat{J}_{i}$ are the Casimir operators of this algebra. We also derived the representation of this algebra which coincides with the previous result obtained in different context\cite{Jakiw}. 

The Hamiltonian in eq.(\ref{Hamil0}) is a new one which becomes a known one when $g=0$. The eigenvalue problem for $g=0$ has been analyzed by several authors\cite{Balazs, Gitman, Argyres}. When $g\neq 0$, 
our representation describes anyonic states with the spin value given by $\sigma=\frac{qg}{c}$. To see this, let us consider eq.(\ref{xj3psi}). The first term on the right hand side corresponds to the orbital angular momentum and the second term ($\sigma$) is the spin of the state. There is no restriction on the value of $\sigma$. The appearence of anyonic states is possible in theories with two-dimensional configuration space. Poincar\'e hyperboloid is an example of two dimensional configuration spaces. It will be interesting to find how $\sigma$ should be restricted when we identify points on Poincar\'e hyperboloid to get a compact Riemann surface with genus greater than 1.


\appendix

\section{Derivation of $\phi(\Lambda,\vec{x})$}

The angle $\phi(\Lambda,\vec{x})$ for a given pair of $(\Lambda,\vec{x})$ is defined by 
\begin{eqnarray} 
	B_{s}^{-1}(\Lambda\vec{x}) \Lambda B_{s}(\vec{x}) = R(\phi(\Lambda, \vec{x})).
\end{eqnarray}
Here $B(\vec{x})$ is the pure boost which transforms $\vec{x}_{s}=(0,0,a)$ to $\vec{x}$. With $\vec{x}=(a\sinh\theta\cos\varphi,a\sinh\theta\sin\varphi,a\cosh\theta)$, we also denote $B(\theta,\varphi)$ instead of $B(\vec{x})$. To derive $\phi(\Lambda,\vec{x})$, we introduce the Poincar\'e disk which is a disk  on the $xy$ plane bounded by a circle of radius $a$ as in Fig. 3. There is a natural one-to-one mapping $\Omega$ from the hyperboloid to the disk as depicted in Fig. 3. 
\begin{figure}[h]
	\centering
    \includegraphics[angle=0, width=0.6 \textwidth]{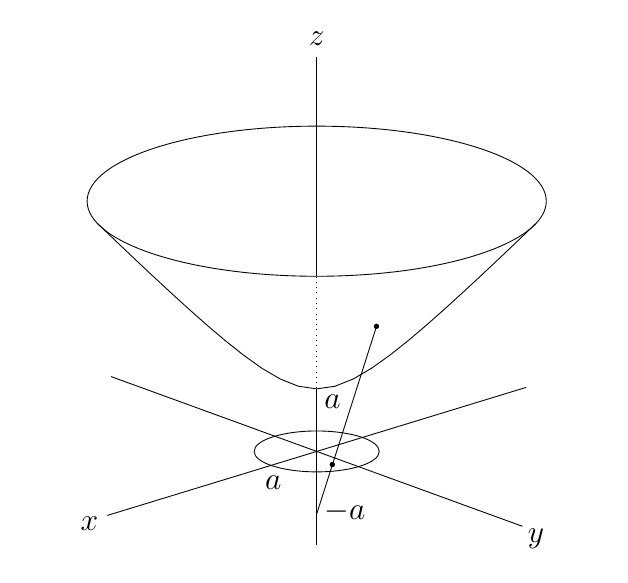}
    \caption{ Poincar\'e disk}
\end{figure} 
A point $(x,y,z)$ on the hyperboloid is mapped to $(\tilde{x}, \tilde{y}, 0)$ on the Poincar\'e disk where 
\begin{eqnarray} 
	\tilde{x} &=& \frac{ax}{z+a},\\
	\tilde{y} &=& \frac{ay}{z+a},
\end{eqnarray}
and substituting $x=a\sinh\theta\cos\phi$, $y=a\sinh\theta\sin\phi$, $z=a\cosh\theta$ the corresponding complex number $\Omega(\vec{x}) = \tilde{x}+i\tilde{y}$ is given by
\begin{eqnarray} 
	\Omega(\vec{x}) = a e^{i\phi} \tanh\frac{\theta}{2}.
\end{eqnarray}
Here we have used the identity
\begin{eqnarray} 
	\frac{\sinh\theta}{1+\cosh\theta} = \tanh\frac{\theta}{2}.
\end{eqnarray}
It is clear that the point corresponding to $\vec{x}_{s} = (0,0,a)$ is the origin of the Poincar\'e disk $\Omega(\vec{x}_{s})=0$. 

We now perform an $x$-directional pure boost $B(\Theta,0)$ on $\vec{x}=(x,y,z)$ and get $\vec{x'}$ given by
\begin{eqnarray} 
	\vec{x'}=(x',y',z')=(\cosh\Theta \,x+\sinh\Theta \,z, y, \sinh\Theta \,x+\cosh\Theta \,z).
\end{eqnarray}
The corresponding point $\Omega(\vec{x'})$ on the Poincare disk is 
\begin{eqnarray} 
	\Omega(\vec{x'}) = \frac{a(\cosh\Theta x+\sinh\Theta z + iy)}{\sinh\Theta x +\cosh\Theta z +a}.
\end{eqnarray} 
Using the identities 
\begin{eqnarray} 
	\cosh\Theta &=& \cosh^{2} \frac{\Theta}{2} +  \sinh^{2} \frac{\Theta}{2} ,\\
	\sinh\Theta &=& 2\cosh \frac{\Theta}{2} \sinh \frac{\Theta}{2},
\end{eqnarray} 
and defining 
\begin{eqnarray} 
	x_{\pm} &=& x\pm i y, \\
	z_{\pm} &=& z\pm a,
\end{eqnarray} 
we get
\begin{eqnarray} 
	\Omega(\vec{x'}) = \frac{a[x_{-}\tanh^{2}\frac{\Theta}{2}+(z_{+}+z_{-})\tanh\frac{\Theta}{2}+x_{+}]}{z_{-}\tanh^{2}\frac{\Theta}{2}+(x_{+}+x_{-})\tanh\frac{\Theta}{2}+z_{+}}.
\end{eqnarray} 
We also have an identity
\begin{eqnarray} 
	x_{+}x_{-} = z_{+}z_{-}.
\end{eqnarray} 
Substituting $x_{-}=z_{+}z_{-}/x_{+}$, we get 
\begin{eqnarray} 
	\Omega(\vec{x'})=\frac{z_{+}z_{-}\tanh^{2}\frac{\Theta}{2} + x_{+}(z_{+}+z_{-})\tanh\frac{\Theta}{2} + x_{+}^{2}}{x_{+}z_{-}\tanh^{2}\frac{\Theta}{2} + (x_{+}^{2}+z_{+}z_{-})\tanh\frac{\Theta}{2} + x_{+} z_{+}}.
\end{eqnarray} 
We see that the denominator and the numerator have a common factor $(z_{-}\tanh\frac{\Theta}{2} +x_{+})$. After factoring out this factor, we obtain
\begin{eqnarray} 
	\frac{\Omega(B(\Theta,0)\vec{x})}{a} = \frac{\frac{\Omega(\vec{x})}{a} + \tanh\frac{\Theta}{2}}{(\tanh\frac{\Theta}{2})\frac{\Omega(\vec{x})}{a}+1 }.
\end{eqnarray} 

Under a pure rotation $R(\Phi)$, $\vec{x}$ transforms to 
\begin{eqnarray} 
	R(\Phi)\vec{x}=(\cos\Phi \,x - \sin\Phi \,y, \sin\Phi \,x + \cos \Phi \,y, z). 
\end{eqnarray} 
Therefore, 
\begin{eqnarray} 
	\frac{\Omega(R(\Phi)\vec{x})}{a} = e^{i\Phi}\frac{\Omega(\vec{x})}{a}.
\end{eqnarray} 

Now for a boost $B(\Theta, \Phi)$ toward a direction $\hat{n}$ which makes an angle $\Phi$ with the $x$-axis, we have the identify
\begin{eqnarray} 
	B(\Theta,\Phi) = R(\Phi)B(\Theta,0) R(-\Phi).
\end{eqnarray} 
Therefore, under $B(\Theta, \Phi)$, $\vec{x}$ transforms to $\vec{x'}$ in three steps and the corresponding $\Omega(\vec{x})$ transforms to $\Omega(\vec{x'})$ as  
\begin{eqnarray}
\frac{\Omega(\vec{x})}{a} &\xrightarrow{R(-\Phi)}& e^{-i\Phi}\frac{\Omega(\vec{x})}{a} \xrightarrow{B(\Theta,0)} \frac{e^{-i\Phi}\frac{\Omega(\vec{x})}{a}+\tanh\frac{\Theta}{2} }{\tanh\frac{\Theta}{2} e^{-i\Phi}\frac{\Omega(\vec{x})}{a} +1}  \nonumber \\ 
&\xrightarrow{R(\Phi)}& \frac{\Omega(\vec{x'})}{a}=e^{i\Phi} \frac{e^{-i\Phi}\frac{\Omega(\vec{x})}{a}+\tanh\frac{\Theta}{2} }{\tanh\frac{\Theta}{2} e^{-i\Phi}\frac{\Omega(\vec{x})}{a} +1}.
\end{eqnarray}
Therefore, we have
\begin{eqnarray} 
	\frac{\Omega(B(\Theta,\Phi)\vec{x})}{a} = \frac{\frac{\Omega(\vec{x})}{a}+\tanh\frac{\Theta}{2}e^{i\Phi} }{(\tanh\frac{\Theta}{2} e^{i\Phi})^{*} \frac{\Omega(\vec{x})}{a} +1}.
\end{eqnarray} 

Finally, under a general Lorentz transformation 
\begin{eqnarray} 
	\Lambda(\phi_{\Lambda}, \theta_{\Lambda}, \varphi_{\Lambda}) = R(\phi_{\Lambda}) B(\theta_{\Lambda},\varphi_{\Lambda}), \label{lamphitheta}
\end{eqnarray}
$\Omega(\vec{x})$ transforms to $\Omega(\Lambda \vec{x})$ with
\begin{eqnarray} 
	\frac{\Omega(\Lambda \vec{x})}{a} = e^{i\phi_{\Lambda}}\frac{\frac{\Omega(\vec{x})}{a}+\tanh\frac{\theta_{\Lambda}}{2}e^{i\varphi_{\Lambda}} }{(\tanh\frac{\theta_{\Lambda}}{2} e^{i\varphi_{\Lambda}})^{*} \frac{\Omega(\vec{x})}{a} +1}. \label{generaleq}
\end{eqnarray} 
This is the formula we will use repeatedly. For example, 
\begin{eqnarray} 
	\frac{\Omega(B(\theta,\varphi)\vec{x}_{s})}{a} = \tanh\frac{\theta}{2}e^{i\varphi}, \label{qqrrtt1}
\end{eqnarray} 
and
\begin{eqnarray} 
	\frac{\Omega(B^{-1}(\theta,\varphi)\vec{x}_{s})}{a} = \frac{\Omega(B(-\theta,\varphi)\vec{x}_{s})}{a} = -\tanh\frac{\theta}{2}e^{i\varphi}. \label{qqrrtt2}
\end{eqnarray}
For the given $\Lambda = R(\phi_{\Lambda}) B(\theta_{\Lambda},\varphi_{\Lambda})$, we introduce $\vec{x}_{\Lambda}$ defined by
\begin{eqnarray} 
	\vec{x}_{\Lambda} = a(\sinh\theta_{\Lambda}\cos\varphi_{\Lambda}, \sinh\theta_{\Lambda}\sin\varphi_{\Lambda}, \cosh\theta_{\Lambda}). 
\end{eqnarray} 
Then, the above formula can be written as 
\begin{eqnarray} 
	\frac{\Omega(\Lambda\vec{x})}{a} = e^{i\phi_{\Lambda}}\frac{\frac{\Omega(\vec{x})}{a}+\frac{\Omega(\vec{x}_{\Lambda})}{a} }{\frac{\Omega^{*}(\vec{x}_{\Lambda})}{a} \frac{\Omega(\vec{x})}{a} +1}.
\end{eqnarray}

In order to derive $\phi(\Lambda,\vec{x})$, we start with 
\begin{eqnarray} 
	\{B^{-1}(\Lambda\vec{x})\Lambda B(\vec{x})\} \vec{x}_{0} = R(\phi(\Lambda,\vec{x})) \vec{x}_{0}		\label{blambdavec}
\end{eqnarray}
for an arbitrary $\vec{x}_{0}$ on the hyperboloid. It is clear that $\phi(\Lambda, \vec{x})$ is independent of $a$ and the choice of $\vec{x}_{0}$. So we put $a=1$. Futhermore we choose $\vec{x}_{0}$ as  
\begin{eqnarray} 
	\vec{x}_{0} = B^{-1}(\vec{x}) \vec{x}_{s}. 
\end{eqnarray}
With this setting and using eq.(\ref{qqrrtt2}), we get 
\begin{eqnarray} 
	\Omega(B^{-1}(\Lambda\vec{x})\Lambda\vec{x}_{s}) = -e^{i\phi(\Lambda,\vec{x})}\Omega(\vec{x}).  \label{blambdavec01}
\end{eqnarray}
The left hand side of this equation is 
\begin{eqnarray} 
	\mathrm{LHS} = \frac{\Omega(\Lambda\vec{x}_{s})- \Omega(\Lambda\vec{x}) }{-\Omega^{*}(\Lambda\vec{x})\Omega(\Lambda\vec{x}_{s})+1} .   
\end{eqnarray}
With $\Lambda = R(\phi_{\Lambda}) B(\vec{x}_{\Lambda}) $, we get 
\begin{eqnarray} 
	\Omega(\Lambda\vec{x}_{s}) = e^{i\phi_{\Lambda}}\Omega(\vec{x}_{\Lambda}).    
\end{eqnarray}
Therefore, eq.(\ref{blambdavec01}) becomes 
\begin{eqnarray} 
\frac{e^{i\phi_{\Lambda}}\Omega(\vec{x}_\Lambda)-\Omega(\Lambda\vec{x})}{ -\Omega^{*}(\Lambda\vec{x})e^{i\phi_{\Lambda}} \Omega(\vec{x}_{\Lambda})+1 } = - e^{i \phi(\Lambda,\vec{x})} \Omega(\vec{x}).	    
\end{eqnarray}
After rearrangement, we get 
\begin{eqnarray} 
	e^{i\phi(\Lambda,\vec{x})} = \frac{e^{i\phi_{\Lambda}}}{\Omega(\vec{x})} \frac{\Omega(\Lambda \vec{x}) e^{-i\phi_{\Lambda}} - \Omega(\vec{x}_{\Lambda} )}{1-(\Omega(\Lambda \vec{x}) e^{-i\phi_{\Lambda}})^{*}\Omega(\vec{x}_{\Lambda})}.   \label{blambdavec02}  
\end{eqnarray}
Substituting
\begin{eqnarray} 
	\Omega(\Lambda\vec{x}) = e^{i\phi_{\Lambda}} \frac{\Omega(\vec{x})+\Omega(\vec{x}_{\Lambda})}{\Omega^{*}(\vec{x}_{\Lambda})\Omega(\vec{x}) + 1 }. 
\end{eqnarray}
into the above equation, we get
\begin{eqnarray} 
	e^{i\phi(\Lambda, \vec{x})} = e^{i\phi_{\Lambda}} \frac{1+\Omega(\vec{x}_{\Lambda})\Omega^{*}(\vec{x})}{1+\Omega^{*}(\vec{x}_{\Lambda})\Omega(\vec{x})}. \label{ephioo1}
\end{eqnarray}
Using 
\begin{eqnarray} 
	\Omega(\vec{x})\Omega^{*}(\vec{x}_{\Lambda})=\tanh\frac{\theta}{2}\tanh\frac{\theta_{\Lambda}}{2}e^{i(\varphi-\varphi_{\Lambda})}, 
\end{eqnarray}
we get
\begin{eqnarray}
	\phi(\Lambda, \vec{x}) = \phi_{\Lambda}+ 2\tan^{-1} \left[ \frac{\tanh\frac{\theta_{\Lambda}}{2}\tanh\frac{\theta_{x}}{2}\sin(\varphi_{\Lambda}-\varphi)}{1+\tanh\frac{\theta_{\Lambda}}{2}\tanh\frac{\theta_{x}}{2}\cos(\varphi_{\Lambda}-\varphi)} \right] . 
\end{eqnarray}

\section{Derivation of $\phi(\Lambda,\Lambda^{-1}\vec{x})$}

To derive $\phi(\Lambda,\Lambda^{-1}\vec{x})$, we substitute $\vec{x}$ in eq.(\ref{ephioo1}) by $\Lambda^{-1}\vec{x}$ and get 
\begin{eqnarray} 
	e^{i\phi(\Lambda, \Lambda^{-1}\vec{x})} = e^{i\phi_{\Lambda}} \frac{1+\Omega(\vec{x}_{\Lambda})\Omega^{*}(\Lambda^{-1}\vec{x})}{1+\Omega^{*}(\vec{x}_{\Lambda})\Omega(\Lambda^{-1}\vec{x})}.  \label{eeeqw}
\end{eqnarray}
Using
\begin{eqnarray}
	\Omega(\Lambda^{-1}\vec{x}) = \frac{e^{-i\phi_{\Lambda}}\Omega(\vec{x})-\Omega(\vec{x}_{\Lambda})}{-\Omega^{*}(\vec{x}_{\Lambda}) e^{-i\phi_{\Lambda}}\Omega(\vec{x})+1}, \label{omegalaminv}
\end{eqnarray} 
we obtain
\begin{eqnarray} 
	e^{i\phi(\Lambda, \Lambda^{-1}\vec{x})} = e^{i\phi_{\Lambda}} \frac{1- e^{-i\phi_{\Lambda}}\Omega(\vec{x})\Omega^{*}(\vec{x}_{\Lambda})}{1- e^{i\phi_{\Lambda}}\Omega^{*}(\vec{x})\Omega(\vec{x}_{\Lambda})}.  \label{eeeqw}
\end{eqnarray}
Using
\begin{eqnarray}
	e^{-i\phi_{\Lambda}} \Omega(\vec{x})\Omega^{*}(\vec{x}_{\Lambda}) = e^{-i(\phi_{\Lambda}+\varphi_{\Lambda}-\varphi)} \tanh\frac{\theta}{2}\tanh\frac{\theta_{\Lambda}}{2},
\end{eqnarray} 
we finally get
\begin{eqnarray}
	\phi(\Lambda,\Lambda^{-1}\vec{x}) = \phi_{\Lambda} + 2 \tan^{-1} \left[  \frac{\tanh\frac{\theta_{\Lambda}}{2}\tanh\frac{\theta}{2}\sin(\phi_{\Lambda}+\varphi_{\Lambda}-\varphi)}{1-\tanh\frac{\theta_{\Lambda}}{2}\tanh\frac{\theta}{2}\cos(\phi_{\Lambda}+\varphi_{\Lambda}-\varphi)}\right].
\end{eqnarray}

\begin{acknowledgments}
This work was supported by Kyungpook National University Research Fund.
\end{acknowledgments}

\end{document}